\pdfoutput=1
\documentclass[PAPER, atlasdraft=false, UKenglish, cernpreprint, texmf, orcidlogo]{atlasdoc}
\usepackage[subcaption]{atlaspackage}
\usepackage{atlasbiblatex}

\usepackage[process]{atlasphysics}

\usepackage[compat=1.1.0]{tikz-feynman}

\usepackage{siunitx}

\graphicspath{{logos/}{figures/}{figs/}}
\usepackage{ANA-TOPQ-2019-14-PAPER-defs}

\AtlasTitle{%
Measurement of the $|V_{cb}|$ element of the CKM matrix in $t\bar{t}$ decays with the ATLAS detector
}

\AtlasAbstract{%
The first measurement of the magnitude of the CKM quark mixing matrix element $|V_{cb}|$ using on-shell $W$-boson decays is presented.
The measurement is made with $t\bar{t}$ events from $pp$ collision data corresponding to an integrated luminosity of $140$ fb$^{-1}$ at a centre-of-mass energy of 13 TeV collected by the ATLAS experiment at the Large Hadron Collider.
A value of:
$$|V_{cb}| = (50\substack{+11 \\ -14})\times10^{-3} %
\nonumber$$
is measured, where the uncertainties have roughly equal contributions from the limited size of the data sample and from systematic effects.
This is consistent with existing measurements made at much lower energy scales in $B$ hadron decays.
While the sensitivity does not yet rival that of previous determinations, this novel approach probes a very different physical situation, namely the rare hadronic $t\rightarrow b\bar{b} c$ decay at high momentum-transfer, rather than semileptonic $B$ decays at low momentum-transfer.
}

\author{The ATLAS Collaboration}

\PreprintIdNumber{CERN-EP-2026-069}

\AtlasJournal{JHEP}
\AtlasJournalRef{\JHEP 07 (2026) 078}
\AtlasDOI{10.1007/JHEP07(2026)078}

\hypersetup{pdftitle={ATLAS document},pdfauthor={The ATLAS Collaboration}}

\begin{document}

\maketitle

\section{Introduction}
\label{sec:intro}
%

% The next lines are included from the .//sections/introduction.tex input file
The Cabibbo-Kobayashi-Maskawa (CKM) matrix~\cite{Cabibbo1963,Kobayashi:1973fv}
is a $3\times3$ unitary matrix which parameterises the couplings of the $W$ gauge bosons to the quarks. Its
elements are among the fundamental parameters of the Standard Model (SM), which can be determined only by experiment. Its $\Vcb$ element was first determined in 1983, following the initial measurement of the typical $B$ hadron lifetime~\cite{b_lifetime_MAC,b_lifetime_MarkII}, which was found to be surprisingly long at ${\cal O}(1$ps). This lifetime, in turn, implied that $\Vcb\sim0.05\ll\Vus$~\cite{PDG_1984}, heralding for the first time the realisation that the CKM matrix was strongly hierarchical~\cite{wolfenstein}.
In the subsequent years, the determination of $\Vcb$ has improved dramatically, both via a sequence of measurements of increasing precision on semileptonic decays of $B$ hadrons, and from improved theoretical understanding~\cite{Aoki:2024FLAG} of the extraction of \Vcb from the measured quantities.
The current PDG average is $\Vcb=(41.1\pm1.2)\times10^{-3}$~\cite{PDG_CKM_Review}, but this masks an unresolved discrepancy of around three standard deviations between the determinations made in inclusive and in exclusive semileptonic $B$ decays. It is thus timely to introduce the novel approach to the measurement of \Vcb reported here, which is based on the method proposed in Ref.~\cite{VcbPaper2019} and uses $W$ bosons produced in the decays of top-quark pairs ($t\bar{t}$).

During LHC Run 2 (2015-2018), ${\cal O}(10^8)$ \ttbar\ pairs were produced and decayed in the ATLAS experiment. In such events, assuming the SM, both top quarks decay via an on-shell $W$ boson, and in this sense, the LHC operates effectively as a $W$ boson factory. While single $W$ events are even more numerous, single-lepton \ttbar events have a big advantage for studying hadronic $W$ decays in ATLAS. Such events allow relatively easy triggering on the charged lepton, which also helps control hadronic backgrounds in offline analysis.
The method used here uses these events, in which one top quark decays into $\bblmnu$ $(\ell\equiv e$, $\mu$ or leptonically decaying $\tau$-lepton), while the other top quark decays hadronically via the \tbWbbc decay.\footnote{Charge-conjugate decays are assumed throughout this paper.} The physical context of the measurement is thus very different from those of traditional approaches using $B$ decays, and the experimental uncertainties obtained are of significantly different origins. It is worth noting that on-shell $W$ bosons were first used to measure a CKM matrix element at LEP 2, but only for the much larger $|V_{cs}|$~\cite{Parkes1998Vcs,ALEPH1999VcsCharmTag,OPAL2000CharmW}.

Traditional measurements of \Vcb~\cite{PDG_Vcb_Review} exploit the decays of $B$ mesons, and are thus made at an effective momentum-transfer squared $(q^2)$ in the region of $\left(\frac{m_b}{2}\right)^2$, where $m_b$ is the mass of the $b$-quark. They use a variety of theoretical techniques in low energy hadronic physics to relate the measured quantities to \Vcb, including lattice gauge theory for the exclusive semileptonic $B$ decays and the heavy quark expansion for the inclusive $B$ decays. These methods each require the introduction of distinct theoretical uncertainties into the respective measurements.
Here, the first measurement of the CKM matrix element $\Vcb$ is presented, made using top quark decays via an on-shell $W$ $(q^2\simeq M_W^2)$. The top quark
decays before it hadronises, and the couplings are probed more directly at the parton level.
Thus, the measurement described here does not rely on the aforementioned theoretical inputs (lattice, heavy quark expansion) and can provide an independent consistency check.
Like other couplings, \Vcb\ is expected to run above the weak scale, albeit quite slowly~\cite{Babu:1987im,CKMRenormalisation}, although
it is expected that such evolution ceases below the weak scale~\cite{PDG_CKM_Review}.
Therefore, if the value of \Vcb\ extracted in this measurement were to differ significantly from that  determined in $B$ decays, the explanation could be either due to new physics in the $\tbbc$ decay or a poorly understood aspect of the low-$q^2$ $B$ decay phenomenology.

\begin{figure}[htbp]
\centering
\includegraphics[width = 0.53\linewidth]{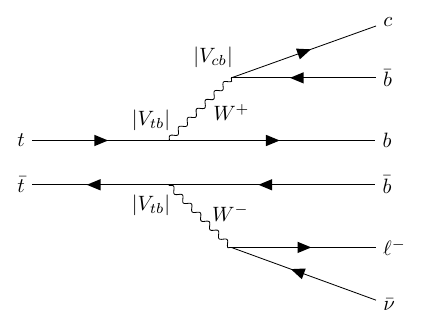}
\caption{Leading-order Feynman diagram describing the $t\bar{t} \rightarrow bW^{+} \, \bar{b}W^{-} \rightarrow b \bar{b}c \, \bar{b} \ell^{-} \bar{\nu}$ decay. Vertices are labelled with the appropriate CKM matrix elements.}
\label{fig:ttbarVcbFeynmanDiagram}
\end{figure}

The measurement determines \Vcb via the ratio of on-shell $\Wcb$ decays to $W^+ \rightarrow c\qbar$, where $q$ is any down-type quark.  The leading-order Feynman diagram for the $t\rightarrow bW^+\rightarrow \bbc$ and leptonic $\tbar$ decay signal process is shown in \Cref{fig:ttbarVcbFeynmanDiagram}.
Taking the ratio of the branching fraction of these decays to that of all similar decays in which the $W$ daughter quark partnering the $c$-quark has any flavour, determines $\Vcb$ via the following relation, valid in the three-generation SM:\footnote{The two top-quarks decay independently. Negligible differences in phase space between the three final states in the denominator are ignored. These are calculable from standard formulae (see e.g.~Ref.~\parencite[Sec.~4.5]{Peskin:1995ev}) and are found to be at the level of $\lesssim 0.3\%$.}
\begin{equation}
\Vcb^2=\frac{\Vcb^2}{\sum_q|V_{cq}|^2}= \frac{{\cal B}(\Wcb)}{{\cal B}(W^+ \rightarrow c\qbar)} =\frac{{\cal B}(\tbbc)}{{\cal B}(\tbqc)}
=\frac{{\cal N}(t\overline{t} \rightarrow \bbc\,\, \bblmnu)}{{\cal N}(t\overline{t} \rightarrow \bqc\,\,\bblmnu)},
\label{eq:ratioDef}
\end{equation}
where $q=d,s,b$, and ${\cal N}$ represents the event yield of the decay mode in question.
This is a challenging measurement due to the small ratio of event yields, $\Vcb^2\simeq1.7\times10^{-3}$~\cite{PDG_CKM_Review}.
Backgrounds and imperfect detector response complicate the extraction of \Vcb\ significantly compared with the simple picture implied by Eq.~\eqref{eq:ratioDef}.
The phenomenological study of the sensitivity of this measurement reported in Ref.~\cite{VcbPaper2019} did not take account of all experimental backgrounds. The analysis presented here applies this method for the first time, using data collected from the ATLAS detector, with a complete treatment of backgrounds and systematic uncertainties.

% End of text imported from the .//sections/introduction.tex input file

%
\section{ATLAS detector}
%
%

% The next lines are included from the .//sections/detector.tex input file
%
\newcommand{\AtlasCoordFootnote}{%
ATLAS uses a right-handed coordinate system with its origin at the nominal interaction point (IP)
in the centre of the detector and the \(z\)-axis along the beam pipe.
The \(x\)-axis points from the IP to the centre of the LHC ring,
and the \(y\)-axis points upwards.
Polar coordinates \((r,\phi)\) are used in the transverse plane,
\(\phi\) being the azimuthal angle around the \(z\)-axis.
The pseudorapidity is defined in terms of the polar angle \(\theta\) as \(\eta = -\ln \tan(\theta/2)\) and is equal to the rapidity
$ y = \frac{1}{2} \ln \left( \frac{E + p_z c}{E - p_z c} \right) $ in the relativistic limit.
Angular distance is measured in units of \(\Delta R \equiv \sqrt{(\Delta y)^{2} + (\Delta\phi)^{2}}\).}

The ATLAS detector~\cite{PERF-2007-01} at the LHC covers nearly the entire solid angle around the collision point.\footnote{\AtlasCoordFootnote}
It consists of an inner tracking detector surrounded by a thin superconducting solenoid, electromagnetic and hadronic calorimeters,
and a muon spectrometer incorporating three large superconducting air-core toroidal magnets.

The inner-detector system (ID) is immersed in a \qty{2}{\tesla} axial magnetic field
and provides charged-particle tracking in the range \(|\eta| < 2.5\).
The high-granularity silicon pixel detector covers the vertex region and typically provides four measurements per track,
the first hit generally being in the insertable B-layer (IBL) installed before Run~2~\cite{ATLAS-TDR-19,PIX-2018-001}.
It is followed by the SemiConductor Tracker (SCT), which usually provides eight measurements per track.
These silicon detectors are complemented by the transition radiation tracker (TRT),
which enables radially extended track reconstruction up to \(|\eta| = 2.0\).
The TRT also provides electron identification information
based on the fraction of hits (typically 30 in total) above a higher energy-deposit threshold corresponding to transition radiation.

The calorimeter system covers the pseudorapidity range $|\eta| < 4.9$. Within the region $|\eta|< 3.2$, electromagnetic calorimetry is provided by barrel and
endcap high-granularity lead/liquid-argon (LAr) calorimeters,
with an additional thin LAr presampler covering $|\eta| < 1.8$
to correct for energy loss in material upstream of the calorimeters.
Hadronic calorimetry is provided by the steel/scintillator-tile calorimeter,
segmented into three barrel structures within $|\eta| < 1.7$, and two copper/LAr hadronic endcap calorimeters.
The solid angle coverage is completed with forward copper/LAr and tungsten/LAr calorimeter modules
optimised for electromagnetic and hadronic energy measurements respectively.

The muon spectrometer (MS) comprises separate trigger and
high-precision tracking chambers measuring the deflection of muons in a magnetic field generated by the superconducting air-core toroidal magnets.
The field integral of the toroids ranges between \num{2.0} and \qty{6.0}{\tesla\metre}
across most of the detector.
Three layers of precision chambers, each consisting of layers of monitored drift tubes, cover the region \(|\eta| < 2.7\),
complemented by cathode-strip chambers in the forward region, where the background is highest.
The muon trigger system covers the range \(|\eta| < 2.4\) with resistive-plate chambers in the barrel, and thin-gap chambers in the endcap regions.

The luminosity is measured mainly by the LUCID--2~\cite{LUCID2} detector that records Cherenkov light produced in the quartz windows of photomultipliers located close to the beampipe.

Events are selected by the first-level trigger system implemented in custom hardware,
followed by selections made by algorithms implemented in software in the high-level trigger~\cite{TRIG-2019-04}.
The first-level trigger accepts events from the \qty{40}{\MHz} bunch crossings at a rate below \qty{100}{\kHz},
which the high-level trigger further reduces in order to record complete events to disk at about \qty{1.25}{\kHz}.

A software suite~\cite{ATL-SOFT-PUB-2021-001} is used in data simulation, in the reconstruction
and analysis of real and simulated data, in detector operations, and in the trigger and data acquisition
systems of the experiment.

% End of text imported from the .//sections/detector.tex input file

%
\section{Data and simulated event samples}
%
%

% The next lines are included from the .//sections/datasets.tex input file
Proton--proton (\pp) collisions at $\rts = 13$~\TeV
collected in 2015--2018 by the ATLAS experiment are analysed~\cite{DAPR-2018-01}.
Only the events where all detector components were operational are selected, resulting in a total data sample corresponding to an integrated luminosity of $140$~\ifb~\cite{DAPR-2021-01}.
Selected events are required to be triggered by one of the single-muon or single-electron triggers which are further described in Section~\ref{sec:objects_and_selection}.

As described in Section~\ref{sec:intro}, the analysis focuses on \ttbar events with a single electron or muon and four jets in the final state. Monte Carlo (MC) simulated event samples are used to estimate the signal and background contributions in such events, and details are provided below in Section~\ref{subsec:ttsamples}.
Heavy-flavour decays are modelled with the \EVTGEN[1.6.0] program~\cite{Lange:2001uf} in all parton shower generators, except the \Sherpa~\cite{Bothmann:2019yzt} generator.
The modelling of backgrounds with subleading contributions to the phase space is described in Section~\ref{subsec:MCminorBG}.
The ATLAS simulation infrastructure~\cite{SOFT-2010-01} is used for all the simulated samples, with the detector response simulated using \GEANT~\cite{Agostinelli:2002hh}.
For the signal \ttbar sample with $\Wcb$ decays, as well as all samples used to evaluate modelling systematic uncertainties, a fast simulation relying on parametrisations of the hadronic showers is used~\cite{ATL-PHYS-PUB-2010-013}.
When evaluating the modelling uncertainties, the fast simulation is used for both the nominal and alternative samples.
All simulated samples are processed with the same reconstruction software as the data.

To account for additional \pp interactions from the same or close-by bunch crossings (pile-up), a set of minimum-bias
interactions generated with \PYTHIA[8.186]~\cite{Sjostrand:2007gs} using the \NNPDF[2.3lo] parton distribution function~(PDF) set~\cite{Ball:2012cx} with the A3 set of tuned parameters (tune)~\cite{ATL-PHYS-PUB-2016-017} is superimposed on the hard scattering events.
Furthermore, the simulated events are reweighted according to the number of additional \pp interactions per bunch-crossing to match the pile-up conditions of each data sample corresponding to the 2015--2018 years of data-taking.
Corrections are applied to simulated events to improve agreement between the data and the simulation samples in the object identification efficiencies, energy scales, and energy resolutions, as described in Section~\ref{sec:objects_and_selection}.

\subsection{\ttbar modelling}
\label{subsec:ttsamples}

All \ttbar MC samples used are simulated using \POWHEGBOX-v2~\cite{Frixione:2007nw,Nason:2004rx,Frixione:2007vw,Alioli:2010xd},
which is based on next-to-leading-order (NLO) QCD matrix element (ME) calculations.
A suite of \ttbar MC samples with various settings are generated to estimate the \ttbar contributions and derive systematic uncertainties for the MC modelling of \ttbar processes.
All \ttbar{} MC samples are normalised
to the next-to-next-to-leading-order (NNLO) cross-section including the resummation of soft gluon emissions at next-to-next-to-leading-logarithmic
(NNLL) accuracy using \TOPpp[2.0]~\cite{Beneke:2011mq,Cacciari:2011hy,Baernreuther:2012ws,Czakon:2012zr,Czakon:2012pz,Czakon:2013goa,Czakon:2011xx}.
The resulting cross-section for the \ttbar process is $\sigma_{\ttbar} = 834^{+21}_{-30}(\text{scale})\pm21(\text{PDF}+\alpha_S)$~pb for $m_t=172.5~\GeV$.

The simulation of \ttbar samples is split according to the $W$ boson decay.
For single lepton events where the hadronic $W$ boson does not decay via $\Wcb$, the simulation uses the simple two-generation quark mixing matrix~\cite{GIMmatrix} with Cabibbo angle $\theta_C$~\cite{Cabibbo1963} given by $\sin{\theta_C}=0.225$.
Events in the orthogonal single lepton sample, denoted by \ttI, are generated by forcing the hadronic top daughter $W$ boson to decay as \Wcb.
The measurement of \Vcb relies on the ratio constructed from these two samples, which differ only in the quark mixing matrix element settings.
All \ttbar events that do not feature \Wcb decay,
including single lepton and dilepton events, are collectively denoted by \ttO.
Both \ttI and \ttO samples are generated in the five-flavour scheme (5FS) using the \POWHEGBOX-v2 ME generator, the \NNPDF[3.0nlo] PDF~\cite{Ball:2014uwa}, the \PYTHIA[8.230] parton shower (PS) generator~\cite{Sjostrand:2014zea} and the A14 underlying event (UE) tune~\cite{ATL-PHYS-PUB-2014-021}.
They also use $\muR=\muF=\sqrt{m_t^2 + \pTtop^2}$ with $m_t=\SI{172.5}{\GeV}$, \hdamp$=1.5m_t$ and \pthard$=0$, where \muR and \muF denote the renormalisation and factorisation scales respectively. The \hdamp parameter controls the first high \pt emission after the Born configuration, while \pthard is a setting that affects the ME-PS matching.
Although the \ttI sample uses the fast simulation for the detector response, the consistency between fast and full simulation has been validated using the \ttO sample, where agreement is observed within the MC statistical uncertainty.

The generated \ttbar events are categorised based on the presence of additional heavy-flavour (HF) hadrons ($b$- and $c$-hadrons), which do not originate from the \ttbar system.
To identify these, a recursive ancestry check is applied to the generator truth record.
The algorithm first counts the total number of HF hadrons in the event. It then subtracts the number of $b$- and $c$-quarks that trace their ancestry directly back to the \ttbar system.
The remaining count defines the number of additional HF hadrons.
Events containing at least one additional $b$-hadron are defined as \ttb.
Events with no additional $b$-hadrons but at least one additional $c$-hadron are defined as \ttc.
Both \ttc and \ttb processes are collectively referred to as \ttHF, while events not classified as \ttHF are referred to as \ttl.

Alternative samples are used to assess systematic uncertainties associated with the modelling of \ttbar in MC samples.
Both the \ttO and \ttI samples are alternatively interfaced with the \HERWIG[7.1.3]~\cite{Bellm:2015jjp,Bahr:2008pv} parton shower generator.
Additional \ttO samples are generated with a variety of $m_t$, \hdamp, \pthard and UE tuning settings.
Furthermore, samples with variations in the final state radiation, $b$-fragmentation, colour reconnection and the recoil to the top quark transverse momentum \pTtop are available to assess systematic uncertainties.
An alternative sample is generated with the four-flavour scheme (4FS) at NLO precision using \POWHEGBOX[Res]~\cite{Jezo:2018yaf} and \OPENLOOPS~\cite{Cascioli:2011va,Denner:2016kdg,Buccioni:2019sur} using the \NNPDF[3.1nnlo] 4FS PDF set~\cite{NNPDF:2017mvq}.
The mass of the $b$-quarks in the ME calculation is set to $m_b=4.95$ \GeV.
The events are then interfaced to \PYTHIA[8.245] with the A14 tune~\cite{ATL-PHYS-PUB-2014-021} and the \NNPDF[2.3lo] PDF set~\cite{Ball:2012cx}.
In this sample, the production of additional $b$-quarks is modelled in the ME, instead of through the PS in the default 5FS sample.
This alternative sample is used to define an additional systematic uncertainty for \ttb.

To serve as pseudo-data for validating the statistical model in~\Cref{sec:results:setup}, an additional \ttbar sample is generated.
This sample is simulated in the 5FS at NLO precision using \SHERPA[2.2.12]~\cite{Bothmann:2019yzt} and \OPENLOOPS, with the \NNPDF[3.0nnlo] PDF set~\cite{Ball:2014uwa}.
The parton showering is performed by \SHERPA based on Catani--Seymour dipole factorisation~\cite{Gleisberg:2008fv,Schumann:2007mg} using the \MEPSatNLO prescription~\cite{Hoeche:2011fd,Catani:2001cc}, with parameters tuned by the \SHERPA authors.

\subsection{Modelling of other backgrounds}
\label{subsec:MCminorBG}
Compared to \ttbar production, other SM backgrounds make a sub-leading contribution to the event rate. They are predicted with MC simulations, unless stated otherwise, and described in the following.

Single-top processes are split into $s$-channel, $t$-channel and $tW$ production. All production modes are modelled with \POWHEGBOX[v2]~\cite{Re:2010bp,Nason:2004rx,Frixione:2007vw,Alioli:2010xd} at NLO, use the \NNPDF[3.0nlo] PDF set, are interfaced to \PYTHIA[8.230] and use the A14 tune for the UE.
To remove overlaps with the \ttbar sample in the $tW$ sample, the diagram removal scheme~\cite{Frixione:2008yi} is used. Similar to the \ttbar samples described in the previous section, a variety of alternative samples, including a $tW$ alternative sample using the diagram subtraction scheme~\cite{Frixione:2008yi,ATL-PHYS-PUB-2016-020}, are generated to assess systematic uncertainties.

The ME, PS and hadronisation for $W$+jets, $Z$+jets and diboson ($WW,WZ,ZZ$) processes is simulated using \Sherpa2.2.11, 2.2.12 or 2.2.14, with the \Sherpa version depending on the boson decay modes. The \NNPDF[3.0nnlo] set of PDFs~\cite{Ball:2014uwa} is used in the ME. These samples are matched with the \SHERPA parton shower~\cite{Schumann:2007mg} %
prescription~\cite{Hoeche:2011fd,Hoeche:2012yf,Catani:2001cc,Hoeche:2009rj}
using the set of tuned parameters developed by the \SHERPA authors. The production of \(V+\)jets is simulated using NLO MEs for up to two partons, and leading-order (LO) matrix elements for up to five partons. %
They are normalised to an NNLO prediction~\cite{Anastasiou:2003ds}.
Samples of diboson final states (\(VV\)) are simulated including off-shell effects and Higgs boson contributions, where appropriate.
Fully leptonic final states and single lepton final states, where one boson decays leptonically and the other hadronically, are simulated using matrix elements at NLO accuracy in QCD for up to one additional parton
and at LO accuracy for up to three additional parton
emissions. Samples for the loop-induced processes \(gg \to VV\) are
simulated using LO-accurate matrix elements for up to one
additional parton emission for both the cases of fully leptonic and single lepton final states.
Finally, a small background comes from \ttbar + $H$, \ttbar + $W$ and \ttbar + $Z$ processes. The production of \ttH events is modelled using the
\POWHEGBOX[v2]~\cite{Frixione:2007nw,Nason:2004rx,Frixione:2007vw,Alioli:2010xd,Hartanto:2015uka}
generator at NLO with the \NNPDF[3.0nlo]~\cite{Ball:2014uwa} PDF set.
The events are interfaced to \PYTHIA[8.230]~\cite{Sjostrand:2014zea}~using
the A14 tune.
The production of \ttV\ events is modelled using the
\MGNLO[2.3.3]~\cite{Alwall:2014hca} generator at NLO with the
\NNPDF[3.0nlo]~\cite{Ball:2014uwa} PDF.
The events are interfaced to \PYTHIA[8.210]~\cite{Sjostrand:2014zea} using the A14 tune~\cite{ATL-PHYS-PUB-2014-021} and the \NNPDF[2.3lo]~\cite{Ball:2014uwa} PDF set.

The multijet background is estimated by using the Matrix Method~\cite{EGAM-2019-01} using fake enriched regions orthogonal to the analysis regions. The total fake and non-prompt lepton efficiency is measured using looser lepton identification and isolation requirements and are used to extrapolate the multijet contribution to the analysis regions.

% End of text imported from the .//sections/datasets.tex input file

%
\section{Event reconstruction and selection}
\label{sec:objects_and_selection}
%

% The next lines are included from the .//sections/objects_and_selection.tex input file
Events are required to have at least one reconstructed \pp interaction vertex with a minimum of two associated tracks with transverse momenta $\pT > 0.5~\GeV$.
The primary vertex is defined as the vertex with the highest sum of squared transverse momenta of associated tracks~\cite{ATL-PHYS-PUB-2015-026}.

Electron candidates are reconstructed from energy deposits in the EM calorimeter matched to a track in the ID~\cite{EGAM-2021-01}.
The track is required to be matched to the primary vertex, where the track longitudinal impact parameter $z_0$, must satisfy $\left|z_0 \sin{\theta}\right| < 0.5$~mm and the ratio of the transverse impact parameter $d_0$ to its uncertainty $\sigma(d_0)$ must satisfy $\left| d_0/\sigma(d_0) \right| < 5$.
Electron candidates are required to be within $\left|\eta\right| < 2.47$, excluding the transition region between the barrel and the endcap calorimeters, $1.37 < \left| \eta \right| < 1.52$.
They must satisfy $\pT > 25~\GeV$ and the \emph{Tight} likelihood identification criteria and \emph{Tight} isolation criteria~\cite{EGAM-2021-01}.

Muon candidates are reconstructed by associating tracks in the ID with tracks or track segments in the MS~\cite{MUON-2018-03}, refined through a global fit that uses hits from both sub-detectors~\cite{PERF-2015-10}.
The track longitudinal and transverse impact parameters must satisfy the requirements $\left|z_0 \sin{\theta}\right| < 0.5$~mm and $\left| d_0/\sigma(d_0) \right| < 3$, respectively.
Muon candidates must satisfy $\pT > 25~\GeV$, $\left|\eta\right| < 2.5$ and the \emph{Medium} identification and \emph{Tight} isolation criteria~\cite{MUON-2018-03}.

Jets are reconstructed using the particle flow algorithm, which combines topological clusters of calorimeter cells with tracking information~\cite{PERF-2015-09,JETM-2018-05}.
These objects are then clustered by the anti-\kt algorithm~\cite{Cacciari:2008gp,Fastjet} with a jet radius parameter $R = 0.4$.
The jet energy scale (JES) is calibrated using both data and simulation, as detailed in Ref.~\cite{JETM-2018-05}. Procedures are adopted to match the jet energy resolution (JER) in simulation to that observed in data~\cite{JETM-2018-05}.
Jets are required to satisfy $\pT > 25~\GeV$ and $\left|\eta\right| < 2.5$.
To suppress jets originating from pile-up interactions, the jet vertex tagger (JVT) multivariate likelihood~\cite{PERF-2014-03} discriminant, using track-based variables, is applied to jets satisfying $\pT < 60~\GeV$ and $\left|\eta\right| < 2.4$, ensuring that the selected jets are matched to the primary vertex.

Jets originating from beauty and charm hadrons ($b$- and $c$-jets) are identified using the DL1r deep neural network algorithm~\cite{FTAG-2019-07}. This algorithm exploits the distinct features of such jets in terms of track impact parameters and displaced vertices reconstructed in the inner detector.
The inputs to the DL1r network include discriminating variables constructed by a recurrent neural network, which exploits the spatial and kinematic correlations between tracks originating from the same $b$- or $c$-hadron. The performance of this tagging algorithm is calibrated in data~\cite{FTAG-2018-01,FTAG-2020-08,FTAG-2019-02}.

Jets are classified using the single DL1r $b$-tagging discriminant output score, $D_b$. A jet is considered $b$-tagged if its score exceeds the threshold corresponding to a 60\% efficiency for correctly identifying $b$-jets in \ttbar events.
A jet is considered $c$-tagged if it is not $b$-tagged but has a score above the threshold corresponding to an 85\% $b$-jet efficiency.
This working point yields a 32\% efficiency for correctly identifying $c$-jets in \ttbar events. The chosen $b$- and $c$-tagging criteria have a rejection factor of 2500 and of 40 on light-jets, respectively. The distribution of the $D_b$ scores for jets from \ttbar events is shown in~\Cref{fig:DbDistribution}.

\begin{figure}[htp]
\centering\includegraphics[width=0.80\linewidth]{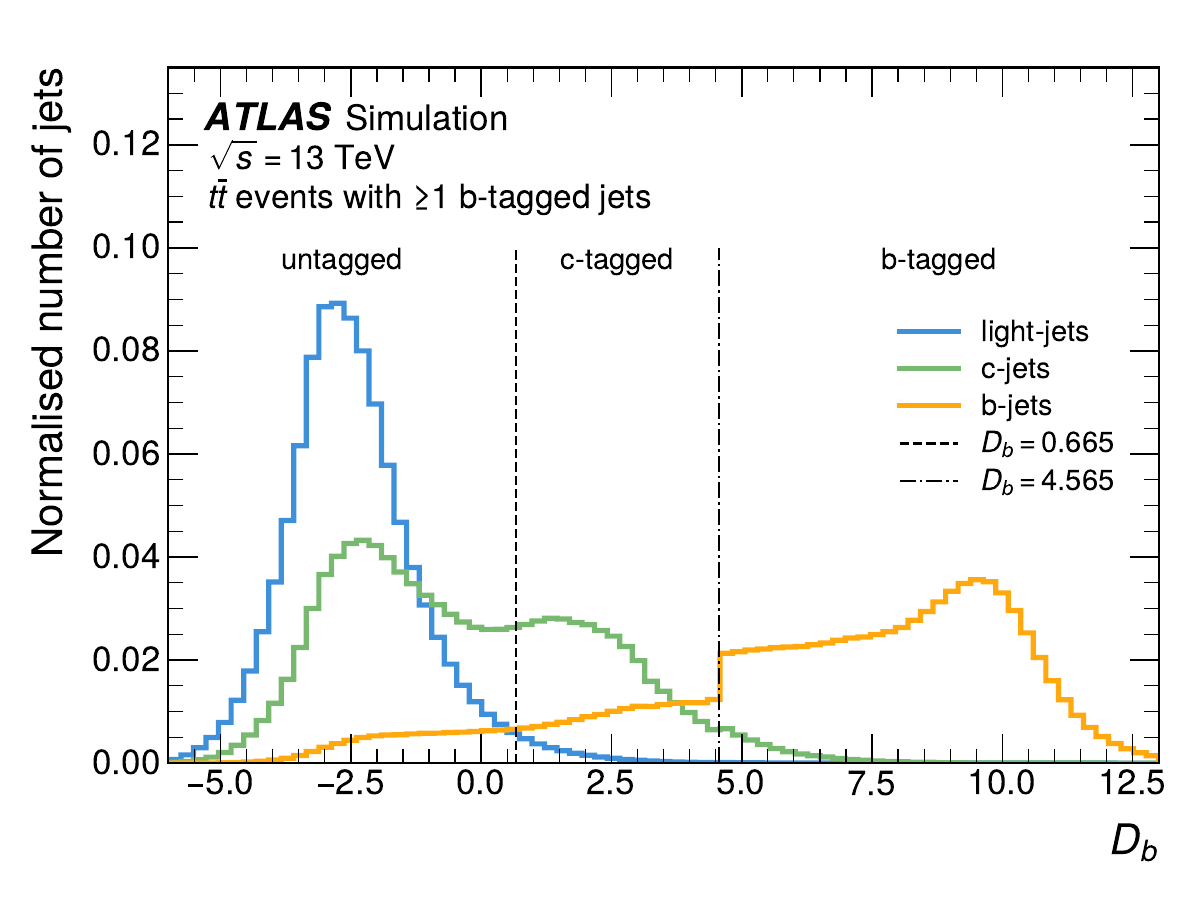}
\caption{Distribution of the $D_b$ scores for jets from \ttbar events after applying the event selection requirements detailed in the text, with the thresholds used to define $c$- and $b$-tagged jets shown. For visualisation purposes, the distributions for each flavour are normalised such that the plotted numbers of true light-, $c$-, and $b$-jets are equal. Since at least one $b$-tagged jet is required in the event selection, there is a step in the distribution at the $b$-tag $D_b$ threshold.}
\label{fig:DbDistribution}
\end{figure}

The missing transverse momentum, $\vec{p}_\mathrm{T}^\mathrm{miss}$,
with magnitude  \met{}, is defined as the negative vector-sum of the transverse momenta of all calibrated objects.
Tracks that are not associated with any of the physics objects but originate from the primary vertex are taken into account as a \met{} soft term~\cite{JETM-2020-03}.

To avoid double-counting the same energy deposits in multiple objects due to reconstruction ambiguities, an overlap removal procedure is applied:
jets within $\Delta R < 0.2$ of an electron are removed;
electrons within $\Delta R < 0.4$ of a remaining jet are removed;
jets with less than three ID tracks within $\Delta R < 0.2$ of a muon are removed to avoid rejecting high-energy muons affected by calorimeter energy loss;
muons within $\Delta R < 0.4$ of remaining jets are removed.

Events are selected using single-electron or single-muon triggers with offline \pT thresholds of 25 \GeV for the 2015 data-taking period, and 27 \GeV for 2016-2018~\cite{TRIG-2018-01,TRIG-2018-05}. Each selected event is required to contain exactly one reconstructed electron or muon. This lepton must also be matched to the object that fired the trigger. Events containing additional electrons or muons with $\pT > 10~\GeV$ satisfying the \emph{Loose} identification criteria~\cite{EGAM-2021-01,MUON-2018-03} are vetoed to suppress backgrounds with multiple leptons. Furthermore, events are required to contain exactly four reconstructed jets with $\pT > 25$ \GeV and $|\eta| < 2.5$. At least two jets must be identified as either $b$- or $c$-tagged, of which at least one must be $b$-tagged.

% End of text imported from the .//sections/objects_and_selection.tex input file

%
\section{Overview of the analysis}
\label{sec:overv}

% The next lines are included from the .//sections/overview.tex input file
As noted above, only events with exactly four reconstructed jets satisfying the kinematic selection are used in the analysis. Jets are tagged uniquely as either $b$-, $c$- or light-flavoured (for all other jets), and the events are divided into tagging categories (TCs), denoted ``$mbnc$'', based on the numbers $m$ of $b$-tagged and $n$ of $c$-tagged jets they contain.
Similarly, production flavours (\TFC s) are defined using the same $mbnc$ notation, based on the truth-level flavour of jets in each event. The truth-level flavour of a jet is assigned by matching the jet to truth-level hadrons within an angular distance of $\Delta R<0.3$ and following the priority order $b>c>\text{light}$~\cite{FTAG-2019-07}.
TC event yields are different from the corresponding \TFC\ event yields, due to jet-flavour mistagging which causes significant numbers of events to migrate between categories.

As suggested by Eq.~\eqref{eq:ratioDef}, sensitivity to \Vcb\ comes primarily from the ratio of the number of events in the $3b1c$ ``signal'' TC to that in the $2b1c$ TC: $\Vcb^2\simeq N(3b1c)/N(2b1c)$.
However, a number of effects complicate this simple picture. First, mistagging of jet flavour means that the signal TC contains additional events which need to be accounted for, according to a suitable calculation involving the numbers of events in other \TFC s and the various jet mistag probabilities which determine the migration between categories. Second, significant numbers of background events enter the signal TC, as a result of physics processes other than pure single lepton weak decays of \ttbar pairs.
In particular, the most significant backgrounds arise from \ttHF events, in which QCD processes generate an extra pair of heavy quarks (\ccbar or \bbbar) in addition to the usual weak decay products of the \ttbar decays, while simultaneously not increasing the number of detected jets. This can happen when, e.g.~quark pairs do not form resolved jets, or when one or more jets fall outside the acceptance, leaving just four jets of uncontrolled flavour.
The rates of these \ttHF processes are often underestimated in simulation, and the approach taken in this analysis is to determine these rates, simultaneously with the measurement of \Vcb.

A final complication of this analysis is that although central ATLAS calibrations of the jet-flavour (mis)tag probabilities (technically, measured weights parameterising data/MC differences) exist for the jet flavours considered, those for true $c$-jets are calibrated using the same channel as the analysis, namely single lepton \ttbar decays~\cite{FTAG-2020-08}, with the \Vcb value fixed. Thus, for this analysis the provided $c$-jet calibration cannot be used directly, and must instead be determined simultaneously with the measurement of \Vcb.
The sensitivity of this channel to the $c$-jet tagging efficiency, $\epsilon_c$, comes directly from the fact that the flavour composition of the vast majority of single lepton \ttbar decays is very well known from the unitarity of the CKM matrix: hadronic $W$ boson decays to two light-flavoured quarks occur with the same rate as those with one light and one c-quark, up to the small contribution given by $\Vcb^2~\sim0.2\%$. Thus, the ratio of the yields of events with these flavours after tagging is directly sensitive to the charm tagging efficiency: $N(2b1c)/N(2b0c)\simeq\epsilon_c/(2-\epsilon_c)$, where the dependence on $\Vcb$ makes only a negligible correction. Furthermore, among genuine single lepton \ttbar decays, the $3b0c$ \TFC\ has negligible rate in the SM as it is suppressed by $|V_{ub}|^2$, and the yield of events seen in the corresponding TC is directly proportional to the $c$-to-$b$ mistagging probability: $N(3b0c)\propto f_{cb}$, allowing the latter to be determined.

From the above discussion, it is clear that the analysis needs to include the numbers of events in several TCs, in addition to the signal TC, $3b1c$. In practice, eight TCs are used, requiring at least two heavy flavour tags, including at least one $b$-tag:
\begin{equation}
1b1c,~1b2c,~1b3c,~2b0c,~2b1c,~2b2c,~3b0c,~3b1c.\nonumber
\end{equation}
The $4b0c$ TC is not used as it does not provide a statistically significant contribution to the analysis.

The \ttHF background events which pass the selection criteria are expected to be incomplete (they have at least six primary partons, with only four jets reconstructed), so that they generally differ kinematically from the primary signal events.
Thus, in order to help separate them from the signal, a neural network (NN), trained to distinguish signal events from such incomplete background events, is deployed. The remaining contribution of the \ttHF background to the signal-enriched regions is constrained using background-enriched regions.
The NN score, $\sigma$, is defined on a scale between 0 and 1, and the events are separated into three distinct regions $(\NNR)$ denoted low, med and high.
The boundaries of the $\NNR$ are obtained by optimising the expected sensitivity to $\Vcb$, resulting in $\sigma_{\text{low}}\leq0.643$, $0.643<\sigma_{\text{med}}<0.736$ and $0.736 \leq \sigma_{\text{high}}$.
The same $\NNR$ boundaries are used across all eight TCs to give $3\times8=24$ ``fit-regions''.
Eventually, the event yields in all 24 regions are fitted using a binned profile-likelihood fit~\cite{Cowan:2010js},
while floating $\Vcb^2$, the total \ttbar\ rate, the rates of \ttc and \ttb, and $\epsilon_c$ and $f_{cb}$ separately in each of the three $\NNR$s.
\Cref{fig:PrefitTCBinnedActivationPlot} illustrates the pre-fit NN score distribution in the signal-enriched $3b1c$ TC, as well as the optimal NN score boundaries.

The following sections describe in more detail the architecture of the NN, the \insitu\ charm tagging, and details of the fit to extract \Vcb.

\begin{figure}[htbp]
\centering
\includegraphics[width=0.8\textwidth]{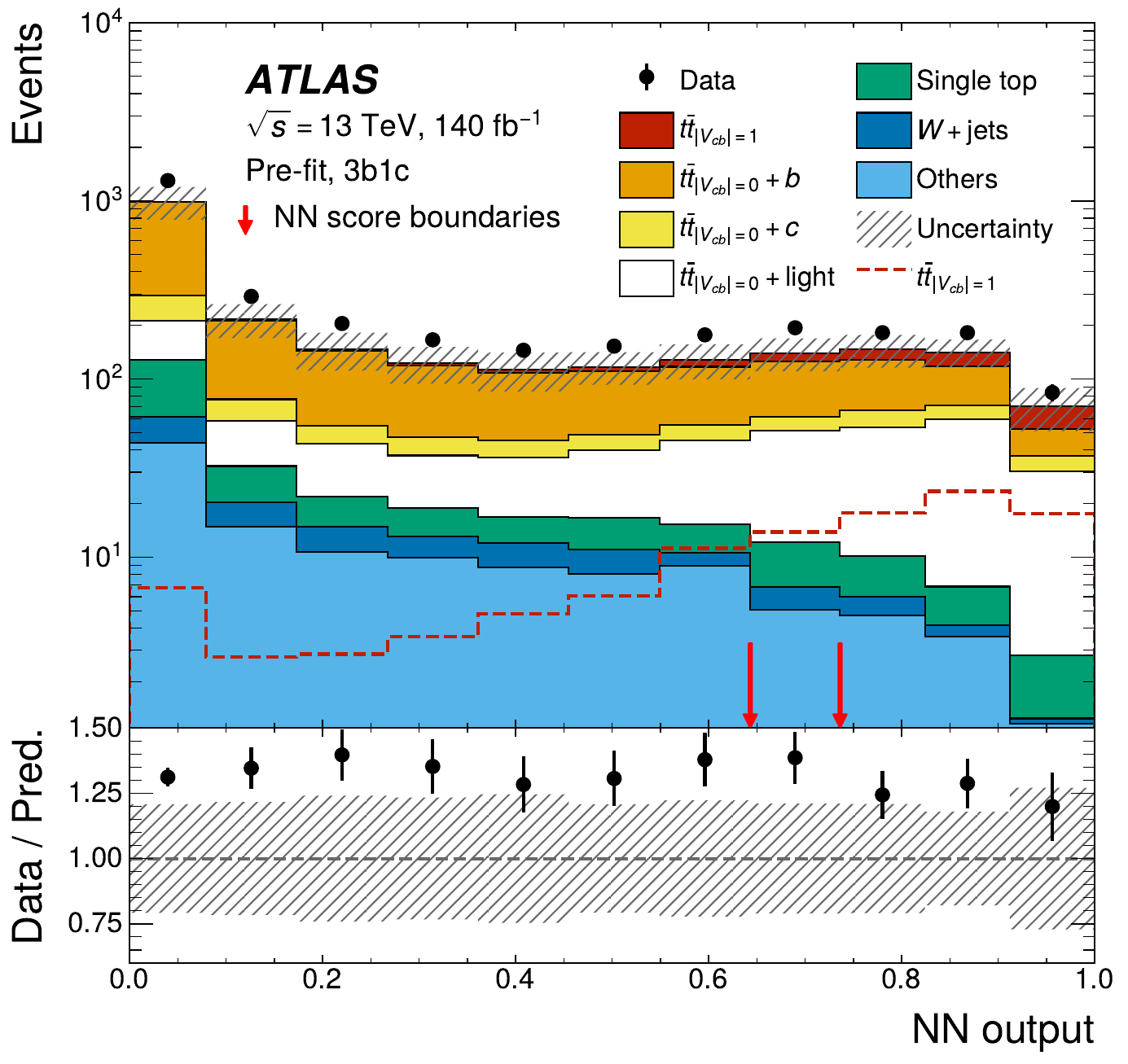}
\caption{Pre-fit distribution of the Neural Network (NN) output score in the $3b1c$ tagging category. The data are shown as closed circles with statistical uncertainties; the prediction is shown as stacked histograms, broken down into the signal \VcbI and background processes. The signal \VcbI process is also overlaid separately for visualisation. The hatched bands represent the total systematic uncertainties. The lower panels show the ratios of the data to the prediction. The arrows indicate the optimised bin boundaries that are used in the profile-likelihood fit. The "Others" category consists of $Z$+jets, diboson, \ttbar + $H/W/Z$ and multijet events.}
\label{fig:PrefitTCBinnedActivationPlot}
\end{figure}

% End of text imported from the .//sections/overview.tex input file

%

\section{Details of the analysis}

\subsection{Neural network configuration}
%

% The next lines are included from the .//sections/neural_network.tex input file
A classifier is designed to discriminate between correctly reconstructed $t\bar{t}$ events and background events, defined in the following paragraph. The classification is based on the output of a NN, which is trained on simulated events and deployed on data and MC. The NN takes a set of reconstructed kinematic inputs from each event, and outputs an activation (score), bound between zero and one, where
background events are assigned a low activation, and signal events a high activation. To avoid biasing the extraction of the $c$-jet tagging probabilities, no flavour-tagging information is used as an input to the NN.

For the NN training, signal events are defined as single lepton $t\bar{t}$ events in simulation, in which the reconstructed lepton and jets can be matched to the corresponding generator-level particle and parton, respectively, within a given angular acceptance. Such events are identified by finding the jet assignment that minimises $\sqrt{\sum_i\Delta R_i^2}$, where $\Delta R_i$ is the angular separation between the $i$'th reconstructed jet and its paired generator-level parton. In accordance with the jet radius used in the jet reconstruction algorithm, a threshold of $\Delta R < 0.4$ is applied to define matched events. In addition, the reconstructed and generator-level leptons are required to be matched within $\Delta R < 0.3$.
Single lepton \ttbar events that fail the matching criteria, as well as simulated events from other processes (including dilepton \ttbar and non-\ttbar), are labelled as background in the training set and weighted according to their respective cross sections.

A set of reconstructed kinematic features is chosen as input, and is split into two distinct input structures. The first is a vector, containing nine features that are independent of the jet assignment (permutation) to parent $W$ and $t$-quarks.
The other nine features in the second input structure do depend on the selected jet permutation, e.g.~the angular distances between the two top candidates, the hadronic $W$ candidate and the lepton.
Additionally, the reconstructed masses of the leptonically and hadronically decaying top-quarks ($m(t_{\text{lep}})$ and $m(t_{\text{had}})$), and the hadronic $W$ boson ($m(W_{\text{had}})$) are included.
There are 12 unique\footnote{The reconstructed quantities used are not sensitive to swapping the assigned $W$ daughter jets.} assignments between four jets and the four partons from a single lepton $t\bar{t}$ decay, each of which yields a unique set of values for the
permutation-dependent features. The second NN input structure is therefore a $12\times9$ matrix, with the rows each representing a fixed jet permutation relative to labels ordered by jet \pT.
The shapes of the input distributions are standardised
to ensure that all inputs have the same absolute scale and width. This is done by applying a Yeo-Johnson power transformation~\cite{powerTransform}, and then normalising each distribution to zero mean and unit standard deviation.

The vector input is processed by a branch of fully connected layers, and the matrix input is processed by a branch of convolutional layers, the latter exploiting the information encoded in the transition between one jet permutation and the next. The outputs of the two branches are %
processed further by a set of fully connected layers.
In addition to background discrimination, the NN is trained to predict the most likely jet permutation as an auxiliary task. This is included as a second output from the NN architecture, and is used for validation plots, but not for the analysis itself.
Two datasets are produced by sorting the full set of simulated events that satisfy the selection criteria into even and odd event numbers. Two NNs with identical architecture are trained, one on each dataset, and are deployed on simulated and observed events in the opposite dataset.
For each dataset, $20\%$ is reserved for validation during the training routine.
Each NN is trained for a maximum of 50 epochs, the internal weights of each model being saved after each epoch. The final trained model is selected as the one with the minimum loss on the validation set. The trained models select signal events with an efficiency of $81\%$ while the corresponding background efficiency is $18\%$, and assigns the jet permutation with an efficiency of $63\%$, when deployed on the validation set.

To verify the importance of the input variables while preserving physical correlations, a local sensitivity analysis~\cite{Simonyan2013} is employed based on the partial derivative of the discrimination output $\sigma$ with respect to each input variable $x_i$.
The importance is quantified by the magnitude of the averaged sensitivity, $|\langle \partial \sigma / \partial x_i \rangle|$, which captures the response to infinitesimal perturbations.\footnote{The perturbation magnitude is set to 1\% of the standard deviation for each input variable.}
As shown in~\Cref{fig:input_sensitivity}, the $W_{\text{had}}$ mass is the most discriminating variable, followed by the softest jet $\pt$ and angular variables; lepton features are retained despite lower sensitivity for their unique information.

\begin{figure}[htbp]
\centering
\includegraphics[width=0.8\linewidth]{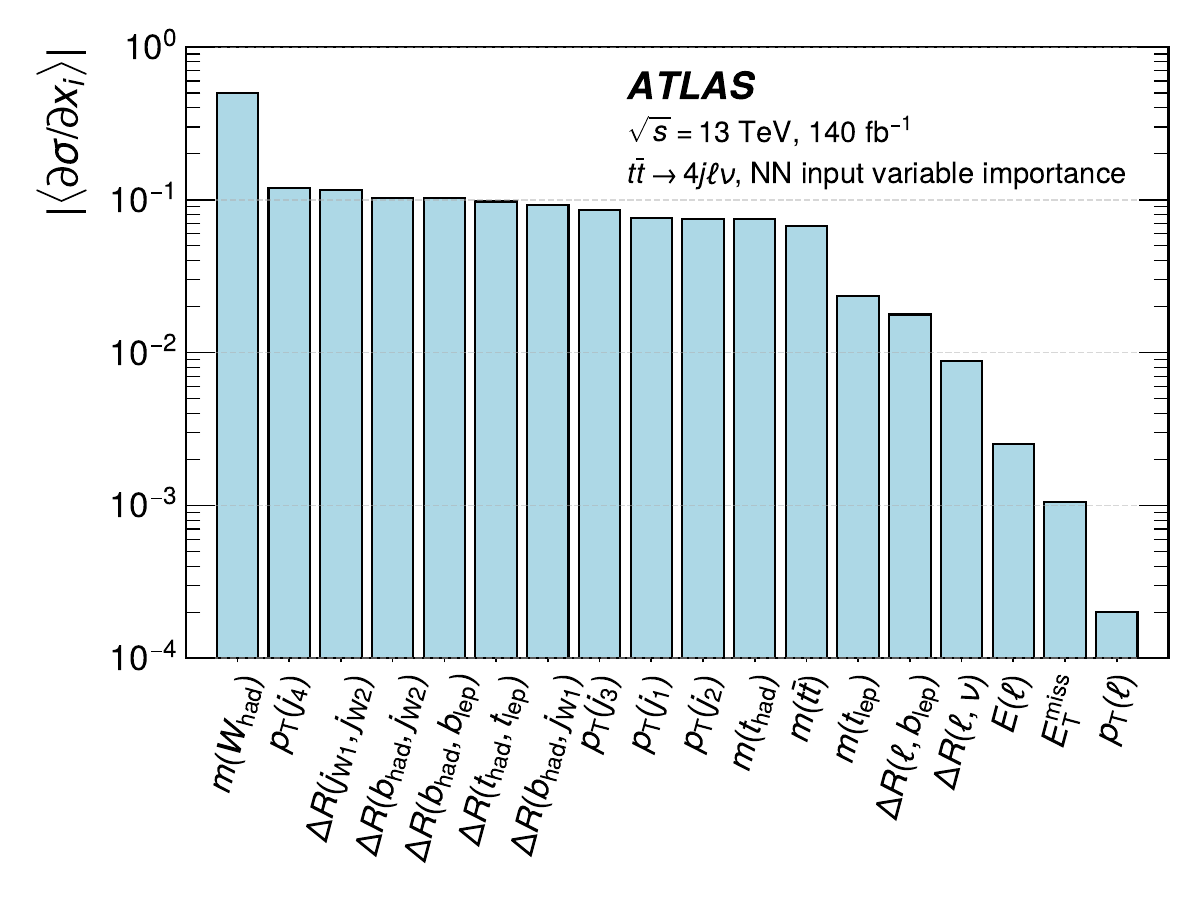}
\caption{Ranking of the neural network input variables by importance, evaluated with data.
The metric is defined as the magnitude of the averaged partial derivative of the discrimination output score with respect to the input. Here, $j_{1,2,3,4}$ refer to the four selected jets in descending order of \pT. The reconstructed neutrino is referred to as $\nu$, whose $p_{x}$ and $p_{y}$ are defined based on \MET, and whose $p_{z}$ is derived using the  $W$ boson invariant mass constraint $m(\ell\nu)=80.4$~\GeV. The hadronic $W$ boson daughter jets are referred to as $j_{W1}$ and $j_{W2}$, in descending order of \pT.}
\label{fig:input_sensitivity}
\end{figure}

% End of text imported from the .//sections/neural_network.tex input file

%

\subsection{\textit{In situ} charm tagging calibration}
%

% The next lines are included from the .//sections/insitu_calibration.tex input file
The $c$-jet tagging efficiencies are calibrated \insitu\ using a data-driven method based on the relative event yields in the different TCs. Jet flavour tagging is described by a matrix of tagging rates that encodes both tagging efficiencies and mistag probabilities between light-, charm-, and bottom-flavoured jets,
\begin{equation}
T =
\begin{pmatrix}
\epsilon_l & f_{lc} & f_{lb} \\
f_{cl} & \epsilon_c & f_{cb} \\
f_{bl} & f_{bc} & \epsilon_b
\end{pmatrix},
\end{equation}
where $\epsilon_x$ denotes the tagging efficiency for jets of true flavour $x$ and $f_{xy}$ the probability for a jet of true flavour $x$ to be tagged as flavour $y$. As each row of $T$ must be normalised to unity, the matrix has six independent parameters, which are chosen to be those in the second and third columns. These tagging rates are defined as kinematically averaged probabilities within the three mutually exclusive {\NNR}s defined in Sec.~\ref{sec:overv}.

Since the standard ATLAS $c$-jet calibration assumes the SM value of \Vcb and relies on the same data sample used in this analysis, the $c$-jet tagging efficiency $\epsilon_c$ and the $c$-to-$b$ mistag rate $f_{cb}$ are instead calibrated directly in the fit. On the other hand,
for $b$- and light-flavoured jets, the tagging rates are corrected by applying the simulation-to-data scale factors and associated uncertainties derived in dedicated calibrations~\cite{FTAG-2018-01,FTAG-2019-02}, to account for the differences between data and simulation.

In a first step, the six independent tagging rates chosen from the $T$ matrix are determined in each \NNR from simulation.
The simulated events include all events from the fifteen possible \TFC s without any tagging preselection.
They are partitioned into a signal process ($k=S$, including only \VcbI) and a background process ($k=B$, including all the other \ttbar and single top-quark processes) to reduce correlations with \Vcb and to improve the sensitivity to the tagging rates.
The determination of the tagging rates is mainly driven by the \VcbO component of the background.
Events are classified into one of the \TFC s according to their truth-level jet flavour composition, and the probability, $\mathcal{P}_{ij}(T)$, for an event in the $i$-th \TFC\ to be reconstructed in the $j$-th TC is computed from combinatorial expressions involving only the tagging rates encoded in $T$. In this calculation, the tagging probabilities for each jet are treated as independent of those for the other jets in the event, an assumption validated by a subsequent closure test.
The probability for an event of process $k$ to be tagged into TC $j$ is then given by
\begin{equation}
p^{k}_{j}(T) = \sum_{i}^{\text{PFs}} t^{k}_{i}\,\mathcal{P}_{ij}(T),
\label{eq:pj}
\end{equation}
where $t^{k}_{i}$ is the fraction of events of process $k$ in the $i$-th \TFC.

A single, shared set of tagging rates is extracted by minimising a total $\chi^{2}$ constructed from the separate contributions of the signal and background processes:
\begin{equation}
\chi^{2} = \sum_{j,k}\left(\frac{N^{k}_{j} - N^{k} \cdot p^{k}_{j}(T)}{\Delta N^{k}_{j}}\right)^{2},
\end{equation}
where $N^{k}_{j}$ is the simulated event yield for process $k$ in TC $j$, $N^{k}$ is the total yield, $p^{k}_{j}(T)$ is the tagging probability given $T$. The uncertainty $\Delta N^{k}_{j}$ is defined as $\sqrt{N^{k}_{j}}$, which models the Poisson fluctuation expected in the real data.
The subscript $j$ runs over the eight considered TCs, and $k$ runs over the signal and background processes, giving sixteen terms in total.
The fit is performed separately in each \NNR to reduce the impact of any kinematic differences between the regions.
Closure tests are performed by applying the extracted tagging rates to~\Cref{eq:pj} to evaluate the predicted event fractions, and then comparing these predictions with the simulated fractions in each TC. Good agreement is observed between the two within the expected statistical fluctuation of the data, as represented by $\Delta N^{k}_{j}$.

The second step is performed as part of the final profile-likelihood fit for \Vcb, in which the $c$-jet tagging efficiency $\epsilon_c$ and $c$-to-$b$ mistag rate $f_{cb}$ are incorporated via floated normalisation factors that correct the event yields in each TC to fit the data.
In this step, all elements of $T$ except $\epsilon_c$ and $f_{cb}$ are fixed to their calibrated values, and distinct tagging rates are used for the low, medium, and high $\NNR$s. The tagging rates, $\epsilon^{\NNR}_c$ and $f^{\NNR}_{cb}$ ($\NNR =$ low, med, high), together with their associated uncertainties, are fitted simultaneously with \Vcb.

The \insitu\ calibration is applied to all \ttbar and single top-quark events simulated with \POWHEG and \PYTHIA. For the other minor processes, a normalisation uncertainty of 30\% per true $c$-jet is assigned.

% End of text imported from the .//sections/insitu_calibration.tex input file

%
\section{Systematic uncertainties}
%
%

% The next lines are included from the .//sections/systematics.tex input file

%
%
%

The measurement of \Vcb is affected by systematic uncertainties arising from experimental sources, the modelling of physics processes, and the finite size of MC samples.
For two-sided uncertainties, where both upward and downward variations are available, the symmetrised uncertainty is calculated as the signed average of the original variations.
In cases of one-sided uncertainties, where only a single alternative scenario is compared against the nominal one, an opposite variation of equal magnitude is assumed.

\subsection{Experimental Uncertainties}

Experimental uncertainties related to the reconstruction of physics objects, pile-up modelling and integrated luminosity are considered.
These uncertainties are derived from dedicated measurements using Run~2 data and are applied to all simulated processes.
This analysis is primarily sensitive to the experimental uncertainties originating from jet energy and light-jet tagging calibrations.

The uncertainty in the combined 2015-2018 integrated luminosity is 0.83\%~\cite{DAPR-2021-01}.
Additionally, the distribution of the average number of interactions per bunch crossing in simulation is rescaled to match that in data.
Uncertainty in this rescaling process is propagated by varying the corresponding reweighting factors.

Uncertainties associated with electrons and muons arise from the trigger, reconstruction, identification, and isolation efficiencies, as well as their momentum scale and resolution.
These are derived from measurements primarily using \Zll ($\ell=e$ or $\mu$) events~\cite{EGAM-2021-01,TRIG-2018-05,MUON-2018-03,TRIG-2018-01,EGAM-2021-02,MUON-2022-01}.

Uncertainties in the JES and JER are evaluated using a series of simulation-based techniques and \insitu\ measurements~\cite{JETM-2018-05,ATL-PHYS-PUB-2022-021}.
These include uncertainty components related to the jet flavour composition, $\eta$-intercalibration, punch-through, single-particle response, calorimeter response to different jet flavours and pile-up effects.
They consist of 30 uncorrelated JES components and 13 uncorrelated JER components.
An uncertainty associated with the JVT efficiency is also included~\cite{PERF-2014-03}.

Uncertainties related to the calibration of the DL1r flavour tagging algorithm for $b$-jets and light-flavoured jets are determined from dedicated measurements using events in the dilepton channel of \ttbar and \Zjets processes, respectively~\cite{FTAG-2018-01,FTAG-2019-02}.
These uncertainties are decomposed into a set of uncorrelated eigenvectors, including 45 components for $b$-jets and 20 components for light-jets.
Additionally, there are four uncertainty components to account for the extrapolation to high-\pT jets, for both flavours.
Uncertainties from Ref.~\cite{FTAG-2020-08} are not applied to $c$-jets, since the flavour tagging for $c$-jets is calibrated within this analysis.

Finally, uncertainties associated with the energy scale and resolution of leptons and jets are propagated to \MET.
Additional uncertainties in \MET arise from momentum scale and resolution uncertainties of the track-based soft term derived from \Zmm events~\cite{JETM-2020-03}.

\subsection{Modelling Uncertainties}

Modelling uncertainties for \ttbar and single top-quark processes are estimated by comparing MC samples generated with baseline settings to those generated with alternative settings.
The dominant systematic uncertainties in this analysis are related to \ttbar modelling, particularly those associated with the modelling of PS and hadronisation.

Uncertainties related to the QCD scale variations in the ME are estimated by varying \muR and \muF in the ME calculation independently by factors 2 and 0.5.
PDF uncertainties are evaluated using the 30 eigenvectors in the PDF4LHC15~\cite{Butterworth:2015oua} prescription, symmetrising the full difference between the nominal PDF4LHC weight and the variation for each eigenvector. Alternative PDF sets corresponding to different \alphas values of 0.117 and 0.119 are also included.
Uncertainties in the initial-state radiation (ISR) modelling are estimated according to the \textsc{Var~3c} set of parameters that are part of the A14 tune~\cite{ATL-PHYS-PUB-2014-021}.
Similarly, the uncertainty related to final-state radiation (FSR) is assessed by varying \muR for final-state parton-shower emissions up and down by a factor of two.
The PS and hadronisation model uncertainty is estimated by comparing the nominal \PYTHIA predictions to those obtained from a sample generated with \POWHEGBOX[v2] using the same settings as the nominal sample for the hard scattering, but interfaced with \HERWIG[7.1.3]~\cite{Bahr:2008pv,Bellm:2015jjp}, using the default set of tuned parameters~\cite{Bellm:2015jjp} and the \MMHT[lo] PDF set~\cite{Harland-Lang:2014zoa}.
The ME-PS matching uncertainty is estimated by comparing the nominal sample to one with the same settings but using \pthard = 1 instead of 0 in \PYTHIA~\cite{ATL-PHYS-PUB-2023-029}.

Since \ttbar is the single most important process in this analysis, more detailed modelling uncertainties are considered.
The uncertainty arising from the limited precision of the fixed-order NLO QCD ME calculation is evaluated by comparing the baseline differential distributions to the NNLO QCD prediction~\cite{Czakon:2017wor}.
The difference is parametrised by reweighting factors derived iteratively using top-quark \pT, \ttbar mass and \ttbar \pT distributions.
The uncertainty related to the modelling of high-\pT radiation is estimated by comparing the nominal sample to one with the \hdamp parameter increased by a factor of 2~\cite{ATL-PHYS-PUB-2016-020}.
The uncertainty related to the top-quark mass is propagated by varying its value by $\pm 0.5$~\GeV.
The uncertainty arising from the ambiguity in the choice of the recoil particle after the first gluon emission~\cite{Brooks:2019xso} in simulating QCD radiation from $b$-quarks in top-quark decay is evaluated, based on alternative samples in which the top-quark participates in the recoil~\cite{Bierlich:2022pfr}.
The uncertainty in the $b$-quark fragmentation function is evaluated by comparing samples generated with $r_b=0.85$, the default value in the A14 tune, with the retuned value $r_b=1.05$ from Ref.~\cite{TOPQ-2017-17}.
An underlying event and multi-parton interaction (MPI) modelling uncertainty is evaluated with alternative samples generated with varied strong coupling constant for MPI and the colour-reconnection range parameter for the proton beam remnants within the \textsc{Var1} parameter set of the A14 tune~\cite{ATL-PHYS-PUB-2014-021}.
The uncertainty related to the modelling of colour reconnection is estimated by comparing samples generated by the "QCD-based" model (CR1) and the "gluon-move" model (CR2) with the baseline sample generated by the "MPI-based" model (CR0). The colour-reconnection uncertainty is then defined by the largest variation for each bin~\cite{TOPQ-2019-01,ATL-PHYS-PUB-2017-008}.

An additional uncertainty is considered for the choice of flavour scheme in the simulation of \ttb events. This is evaluated by comparing the 5FS baseline sample, assuming a massless $b$-quark, with a 4FS alternative sample, generated with a $b$-quark mass of 4.95~\GeV, which is consistent with the $b$-quark mass in the top-quark decays.

The \ttbar uncertainties are treated as uncorrelated between the single lepton and dilepton processes, as at least two of the selected jets have to come from ISR or FSR in dilepton events.
While most modelling uncertainties are correlated between \ttO and \ttI, the \ttbar PS and hadronisation uncertainty is treated as uncorrelated to account for their different heavy-flavour compositions.
Furthermore, the PS and hadronisation uncertainty is split into \ttl, \ttc, and \ttb components, because $c$- and $b$- jets are more likely to be produced via gluon splitting in \ttc and \ttb processes, respectively.
These decorrelation treatments are designed to model the distinct effects independently and prevent unphysical propagation of constraints.
Ultimately, the best-fit \Vcb and its uncertainty are not found to depend significantly on these treatments.

For the dominant background process \ttOl, the uncertainties related to the PS and hadronisation, FSR, and colour reconnection are decomposed into TC and NN components.
The TC component captures event migration between TCs, while the NN component captures event migration between NN score bins within each TC.
This treatment prevents the fit from propagating constraints from a sensitive component to an insensitive one, which would otherwise underestimate the total uncertainty.
To further reduce the constraint on the NN component of PS and hadronisation uncertainty in \ttOl, the original variation is 50\% decorrelated across the NN score bins by splitting it into one correlated and three uncorrelated components, each scaled by $1/\sqrt{2}$.
This treatment significantly relaxes the constraints associated with this source of the uncertainty while having a minor impact on the \Vcb sensitivity.

An extra uncertainty in the calculation of the interference between $tW$ and \ttbar processes is considered for single top-quark modelling. This uncertainty is estimated by comparing the baseline $tW$ sample simulated using diagram-removal, to an alternative sample using diagram-subtraction~\cite{Frixione:2008yi,ATL-PHYS-PUB-2016-020} with otherwise identical settings.

The \ttbar normalisation is determined in the fit to data.
A cross-section uncertainty of 5.4\% is applied to single top-quark processes~\cite{Kidonakis:2021vob}.
For the remaining processes, including \Wjets, \Zjets, diboson and $\ttbar+X$, an overall uncertainty of 45-50\% is applied~\cite{ATL-PHYS-PUB-2017-006,ATL-PHYS-PUB-2017-005}. An overall 100\% uncertainty is applied to the small multijet background.
These uncertainties have a negligible effect on the final results.

% End of text imported from the .//sections/systematics.tex input file

%
\section{Statistical analysis and results}
%
%

% The next lines are included from the .//sections/statistical_analysis_and_results.tex input file
\subsection{Fit setup}
\label{sec:results:setup}

The value of \Vcb is extracted using a binned profile-likelihood fit~\cite{Cowan:2010js} implemented within the \textsc{TRExFitter} framework~\cite{TRExFitter}.
A normalisation factor, $\mu_{|V_{cb}|}$, defined as the ratio of the post-fit value to the pre-fit one, is used throughout this section. The pre-fit value of \Vcb is set to a nominal value of $0.04173$, used by the event generators. In addition to $\mu_{|V_{cb}|}$, the overall normalisations of the \ttbar, \ttbar+$c$, and \ttbar+$b$ components are treated as free parameters. The $c$-jet tagging efficiencies and mistag rates are also treated as free parameters and are independent in the three NN score regions.
Systematic uncertainties are incorporated as nuisance parameters that affect the predicted event yields through bin-by-bin variations. Most nuisance parameters are constrained by Gaussian probability density functions. The nuisance parameters related to the statistical uncertainty of the simulated samples are constrained by gamma distributions following the Beeston-Barlow technique~\cite{Barlow:1993dm}.
The fit is performed simultaneously across 24 bins, comprising eight TCs for each of the three NN score regions.

The likelihood function is the product of Poisson probability terms over all analysis bins, multiplied by constraint terms for the nuisance parameters, and is defined as
\begin{equation}\mathcal{L}(\mathrm{data}\mid \mu_{|V_{cb}|}, \boldsymbol{\mu}, \boldsymbol{\theta}) =
\prod_{k \in \mathrm{bins}}\mathrm{Pois}\left(n_k \mid \nu_k(\mu_{|V_{cb}|}, \boldsymbol{\mu}, \boldsymbol{\theta})\right)\;\times\;\prod_{i}\mathrm{Gauss}\left(\theta_i \mid \theta_{i,0}, 1\right),\end{equation}
where $n_k$ denotes the observed number of events in bin $k$, and $\nu_k$ is the corresponding expected yield, which depends on the signal strength parameter $\mu_{|V_{cb}|}$, the set of normalisation factors $\boldsymbol{\mu}$, and the nuisance parameters $\boldsymbol{\theta}$.
All nuisance parameters, except those accounting for the limited size of the simulated samples, are constrained by Gaussian terms.

Systematic uncertainties with an impact of less than 0.5\% on the yield of a given sample in a given bin are omitted for that sample and bin to improve numerical stability and computational performance.
It was verified that variations of this threshold have a negligible impact on the fit results.

Table~\ref{tab:norm_factor_table} summarises the impact of the fit parameters on the different simulated samples.
A common normalisation factor is applied to all \ttbar samples, while additional scaling parameters control the relative contributions of the heavy-flavour components.
With this parametrisation, the value of \Vcb is directly extracted from the fit through its effect on the \ttI normalisation.

The model is fit to pseudo-data to validate its robustness.
The pseudo-data is constructed from the total prediction, where the contribution of the nominal \ttbar sample is replaced by the \SHERPA \ttbar prediction.
The results indicate that the fit to the pseudo-data can recover the inserted values of $\mu_{\Vcb}$ and the other free parameters within the considered modelling uncertainty.

\renewcommand{\arraystretch}{1.75}
\begin{table}[htbp]
\centering
\caption{Overview of the normalisation and tagging parameters used in the fit and their effect on the simulated samples.}
\begin{tabular}{l l}
\hline
Parameter & Affected samples / quantities \\
\hline
$\mu_{|V_{cb}|}$
& \ttI~(all), yield scaled with $\mu_{|V_{cb}|}^2$ \\

$\mu_{\ttbar}$
& \ttI~(all), \ttO~(all)\\

$\mu_{\ttbar+c}$
& \ttIc, \ttOc \\

$\mu_{\ttbar+b}$
& \ttIb, \ttOb \\

$\epsilon_c^{\text{low,med,high}}$
& $c$-jet tagging efficiency in low, med, and high regions \\

$f_{cb}^{\text{low,med,high}}$
& $c$-to-$b$ mistag rate in low, med, and high regions \\
\hline
\end{tabular}
\label{tab:norm_factor_table}
\end{table}

\subsection{Fit results}
The value of \Vcb extracted from the fit is
\begin{align*}
\Vcb &= (50\substack{+7 \\ -9}~(\text{stat.}) \, \substack{+9 \\ -10}~(\text{syst.}))\times10^{-3}\\
&= (50\substack{+11 \\ -14})\times10^{-3}.
\end{align*}
This corresponds to $\Vcb^2 = (2.5 \pm 1.2)\times10^{-3}$, which scales the decay rate of \Wcb.
The observed (expected) significance relative to the hypothesis in which $\Vcb=0$ is $2.0\sigma$ ($1.9\sigma$).
The value of \Vcb measured in this analysis is consistent, within one standard deviation, with both inclusive and exclusive measurements from $B$ decays, and with the current world average of $\Vcb=(41.1\pm1.2)\times10^{-3}$~\cite{PDG_CKM_Review}.
This consistency supports the Standard Model descriptions of both the low-$q^2$ phenomenology of $B$ decays, and of the much higher $q^2$ decays \tbbc which this analysis probes.

The consistency of the high-$q^2$ measurement of $\Vcb$ with the three generation Standard Model allows the measurement to be re-interpreted in terms of the branching fraction ${\cal B}(\tbbc)$.
Eq.~\eqref{eq:ratioDef} reduces to
\begin{equation}
\Vcb^2
=\frac{{\cal B}(\tbbc)}{{\cal B}(\tbqc)},
\end{equation}
while the unitarity of the CKM matrix gives
${\cal B}(\tbqc)=\tfrac{1}{2}{\cal B}(\tbqq)$
exactly.\footnote{There are corrections of order $\leq 10^{-10}$ to these branching fraction formulae from loop-mediated FCNCs in the SM~\cite{Eilam:1990zc, Mele:1998ag}.} Thus
\begin{equation}{\cal B}(\tbbc)=\tfrac{1}{2}|V_{cb}|^2{\cal B}(\tbqq),
\end{equation}
from which the measured branching fraction,
${\cal B}(\tbqq)=(66.5\pm1.4)\%$~\cite{TOPQ-2014-01}, gives
\begin{equation}{\cal B}(\tbbc)=(8\pm4)\times10^{-4}.
\end{equation}

The post-fit data and predicted yields in all analysis bins are shown in~\Cref{fig:postfit_all_regions}.
\begin{figure}
\centering
\includegraphics[width=0.97\linewidth]{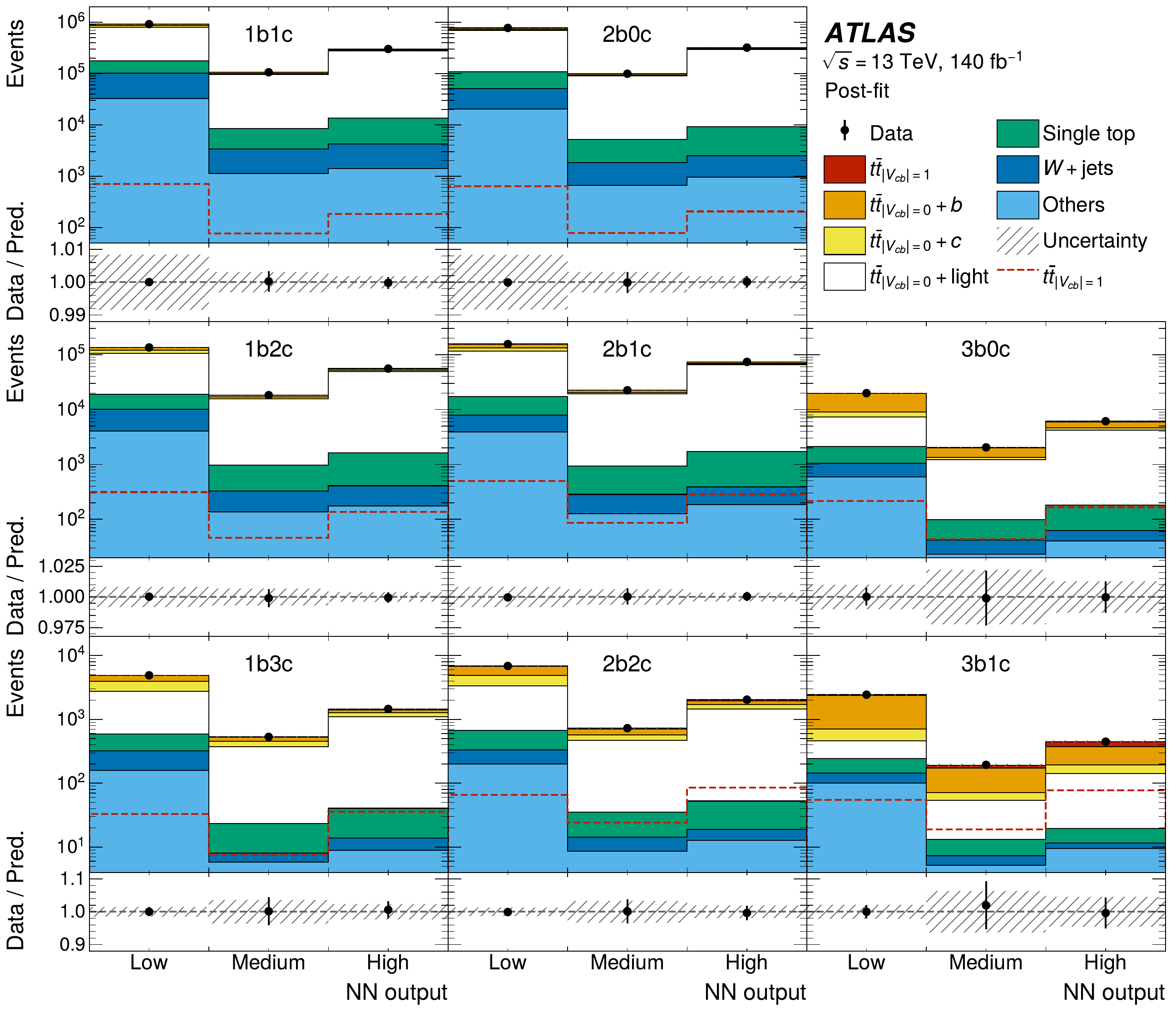}
\caption{Post-fit distributions of the data and predicted event yields across all bins used in the fit. Each panel in the grid represents a tagging category ($1b1c$-$3b1c$), showing the event yields in the three Neural Network (NN) output score bins (Low, Medium, High). The grid is organised such that regions in the same row contain the same total number of heavy-flavour ($b$ or $c$) jets and share the same $y$-axis scales, while regions in the same column contain the same number of $b$-jets. The data are shown as closed circles with statistical uncertainties; the prediction is shown as stacked histograms, broken down into the signal \VcbI and background processes. The signal \VcbI process is also separately overlaid for visualisation. The hatched bands represent the total systematic uncertainties. The lower panels show the ratios of the data to the predictions.}
\label{fig:postfit_all_regions}
\end{figure}
\begin{figure}
\centering
\includegraphics[width=0.8\linewidth]{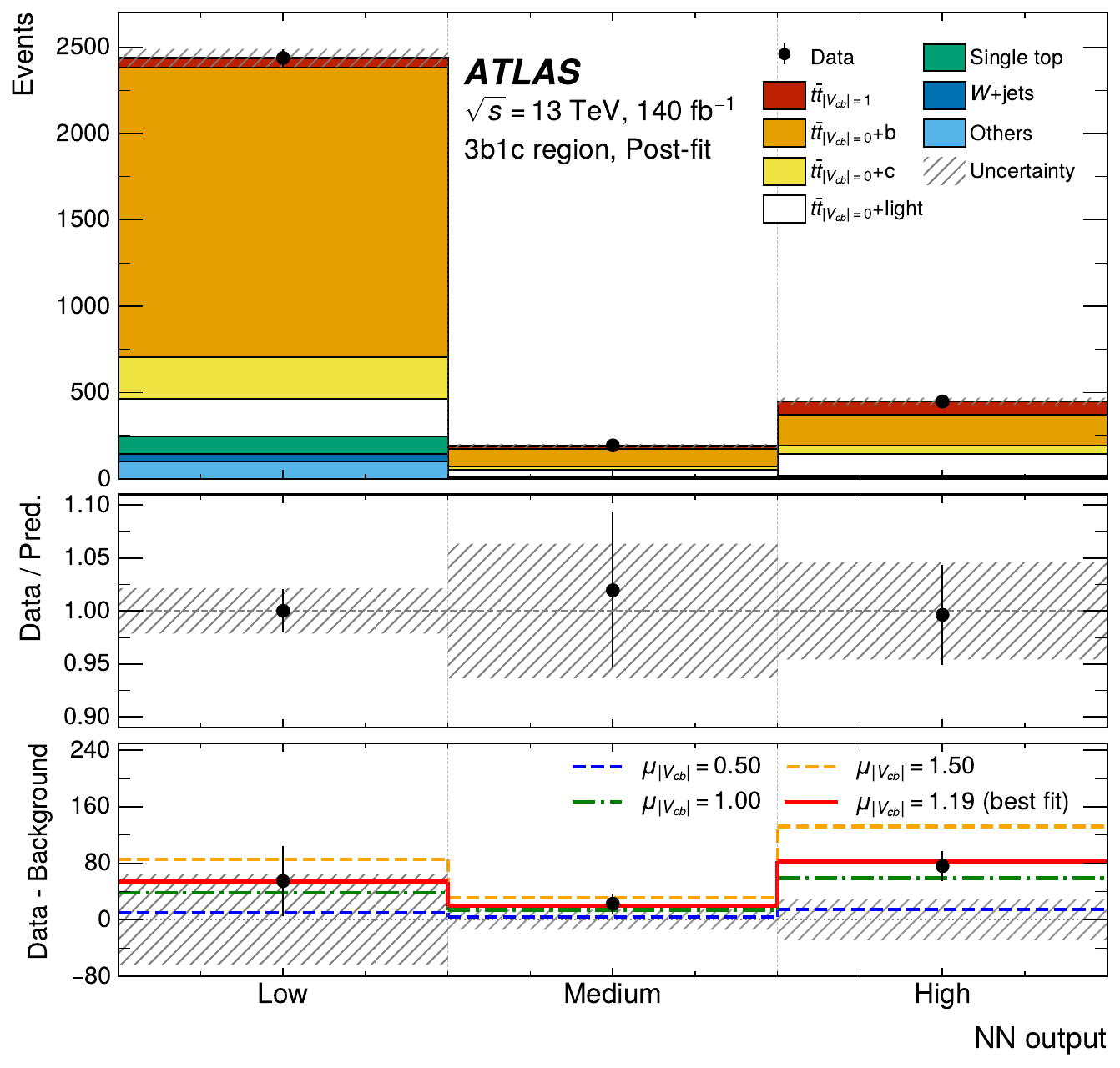}
\caption{Post-fit distribution of the NN output in the 3b1c region. The upper panel shows the data yield in each bin, and the fitted background and signal MC samples. The middle panel shows the ratio between data and MC. The shaded region marks the total systematic uncertainty, while the error bar on the data symbols represents the statistical uncertainty. The bottom panel shows the data with the background (all samples, excluding \ttI) subtracted, with the lines showing the expected number of \ttI events at different values of $\mu_{|V_{cb}|}$, together with the best fit value.}
\label{fig:postfit_3b1c}
\end{figure}
The fit has many free parameters and the data are well described by the fitted model in all regions, with an excellent goodness-of-fit, evaluated by
a saturated model~\cite{Cowan:2010js}.
\Cref{fig:postfit_3b1c} shows the post-fit distribution in the 3b1c region. The bottom panel displays the data after subtraction of the predicted background contributions and compares it to the expected signal yields for different values of the \Vcb normalisation factor $\mu_{|V_{cb}|}$, including the best-fit result.

Table~\ref{tab:Res:NF}
\begin{table}[htbp]
\renewcommand{\arraystretch}{1.3}
\centering
\caption{Best fit values of the freely floating parameters extracted from the fit. The values of $\epsilon_c$ and $f_{cb}$ are shown in the three different NN score regions.}
\begin{tabular}{cc|cccc}
\toprule
Parameter & Value & Parameter &  & Value &  \\
\midrule
$\mu_{|V_{cb}|}$ & $1.19\substack{+0.27 \\ -0.35}$ &  & \underline{NN Low} & \underline{NN Medium} & \underline{NN High} \\
$\mu_{t\bar{t}}$  & \quad$0.99 \pm 0.03$ & $\epsilon_c$ & $0.35 \pm 0.02$ & $0.34 \pm 0.02$ & $0.33\substack{+0.02 \\ -0.01}$  \\
$\mu_{t\bar{t}+c}$ & $1.60\substack{+0.64 \\ -0.60}$ & $f_{cb}$ ($10^{-3}$) & $25 \pm 6$ & $26 \pm 3 $ & $26 \pm 2$ \\
$\mu_{t\bar{t}+b}$ & $1.45\substack{+0.54 \\ -0.34}$  &  &  &  &  \\
\bottomrule
\end{tabular}
\label{tab:Res:NF}
\end{table}
summarises the central values and uncertainties of the free-floating parameters of the fit. The overall \ttbar normalisation is found to be in good agreement with the SM prediction. The normalisations of the \ttc and \ttb components are measured to be larger than predicted by simulation, consistent with earlier measurements of the same processes~\cite{TOPQ-2019-03,TOPQ-2021-26}. The uncertainties in these normalisations are larger in this analysis than in the dedicated measurements. This is because the present analysis uses only the four-jet topology, whereas the dedicated measurements use events with higher jet multiplicities, which provide better sensitivity to these processes. An additional validation of these normalisations is performed in regions with 5-6 jets containing at least one additional c- or b-jet, where good agreement is found between the data and the scaled predictions.
The fitted c-jet tagging efficiency and mistag rates are both consistent with a constant value across the neural network score regions and with prior ATLAS calibrations~\cite{FTAG-2020-08}.

Table~\ref{tab:Res:VcbUnc} shows the contribution of the different sources of systematic uncertainty to the total uncertainty in \Vcb. These contributions are estimated by using a toy MC approach, in which 5000 pseudo-experiments are generated for each uncertainty group. The resulting uncertainty breakdown is found to be consistent with the total uncertainty when summed in quadrature. As a cross-check, the impact of each systematic uncertainty is also estimated by using the covariance between $\mu_{|V_{cb}|}$ and the corresponding nuisance parameter~\cite{pinto2024uncertaintycomponentsprofilelikelihood}. This alternative method yields a similar ranking of the dominant uncertainties. The impact in $\mu_{|V_{cb}|}$ for the 15 highest ranked systematic uncertainties is shown in Figure \ref{fig:Ranking_covariance}. The most important uncertainties are those associated with variations of the parton shower model between \PYTHIA and \HERWIG, which lead to large uncertainties for the \ttIl, \ttOl, \ttOc, and \ttOb samples.

{
\renewcommand{\arraystretch}{1.0}
\begin{table}[htbp]
\centering
\caption{Breakdown of the contribution from each source to the total uncertainty in \Vcb. The uncertainty components are estimated with a toy MC method, where 5000 toy samples are generated for each uncertainty group.
The leading contributions to the `Others' category in \ttbar modelling are from final-state radiation, matrix-element and parton-shower (PS) matching, and colour reconnection.
The contribution from {\insitu} $c$-jet calibration cannot be measured using the same approach. Instead, it is incorporated into the statistical uncertainty along with the contributions from other freely floating parameters.}
\begin{tabular}{lc}
\toprule
Uncertainty in \Vcb & Up/Down [$10^{-3}$]  \\
\midrule
Modelling of \ttbar & $8/9$ \\
~$\hookrightarrow$~PS and hadronisation & $5/6$ \\
~$\hookrightarrow$~Others & $5/6$ \\
Jet flavour tagging calibration & $2.9/3.1$ \\
~$\hookrightarrow$~Light-jet & $2.5/2.8$ \\
~$\hookrightarrow$~$b$-jet & $1.2/1.3$ \\
Simulation statistical uncertainty & $2.6/2.7$ \\
Jet calibration & $2.4$ \\
Modelling of non-\ttbar processes & $1.6$ \\
Others & $0.8$ \\
\midrule
Systematic uncertainty & $9/10$ \\
Statistical uncertainty & $7/9$ \\
\midrule
Total uncertainty & $11/14$ \\
\bottomrule
\end{tabular}
\label{tab:Res:VcbUnc}
\end{table}
}

\begin{figure}[htbp]
\centering
\includegraphics[width=0.8\textwidth]{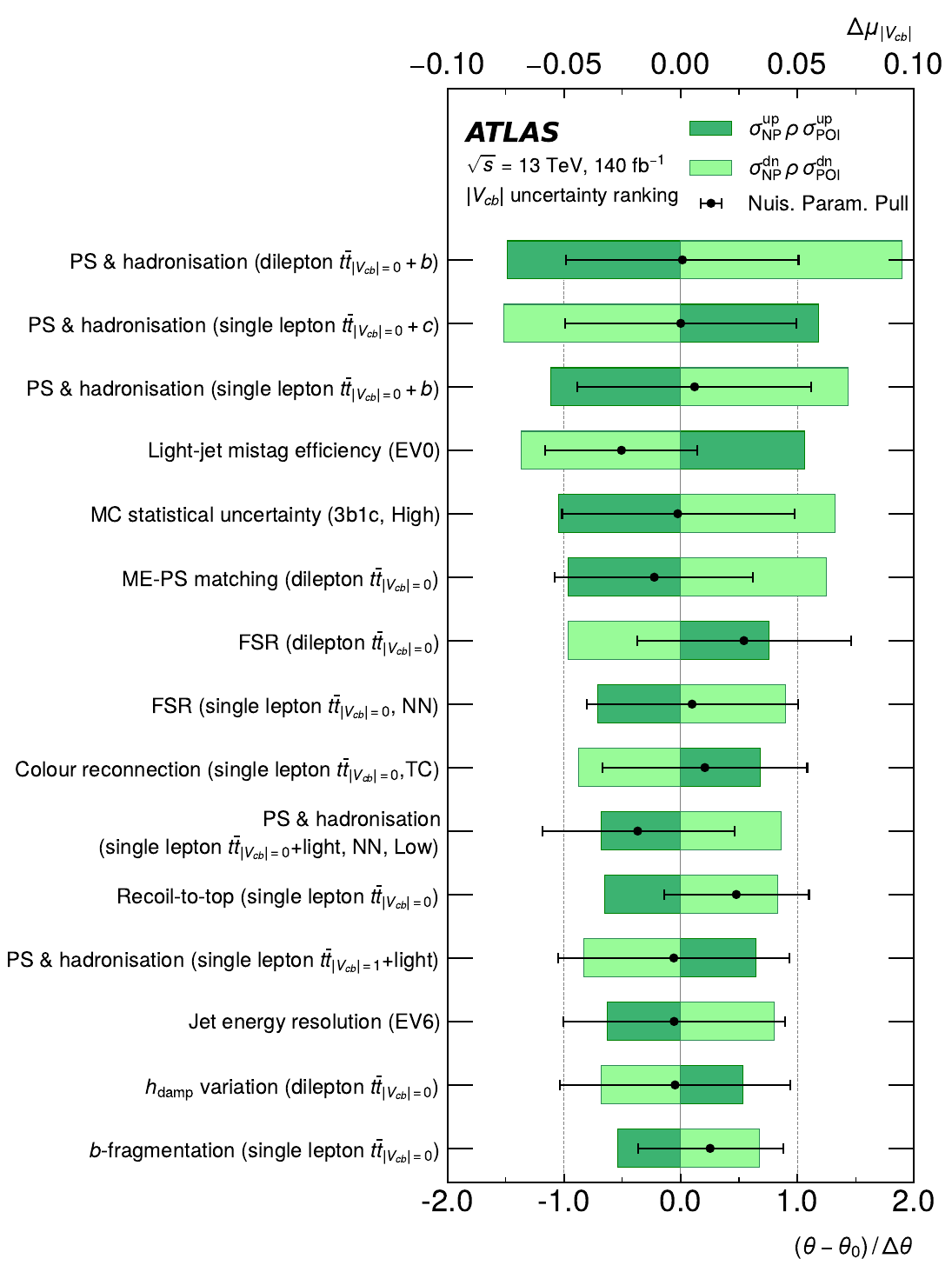}
\caption{Ranking of the fifteen most significant impacts of individual systematic uncertainties in $\mu_{\Vcb}$ using the covariance method. The dots shows the post-fit value of each nuisance parameter, and the associated band shows the post-fit uncertainty of the parameter. Uncertainties that are decorrelated between samples or regions are labelled with the corresponding sample or region in parentheses. The light-jet mistag efficiency and jet energy resolution uncertainties are split into components; the parentheses indicate the most impactful component.}
\label{fig:Ranking_covariance}
\end{figure}

% End of text imported from the .//sections/statistical_analysis_and_results.tex input file

%
%
\FloatBarrier

\section{Conclusion}
%
%

% The next lines are included from the .//sections/conclusions.tex input file
The first measurement of the CKM matrix element \Vcb at the weak scale is presented. The measurement uses single lepton \ttbar events with exactly four jets, taken from $140$~fb$^{-1}$ of $pp$ collision data at $\sqrt{s} = 13$~\TeV collected with the ATLAS detector at the LHC. The measurement determines \Vcb, via the ratio of on-shell $\Wcb$ decays to $W^+ \rightarrow c\qbar$, where $q$ is any down-type quark. The normalisations of \ttbar processes with additional heavy quarks, together with the charm-jet tagging efficiency and mistag rates are determined simultaneously.

The measured value
\begin{equation}
\Vcb = \left(50\substack{+11 \\ -14}\right)\times10^{-3}
\end{equation}
is found to be consistent, within one standard deviation, with earlier measurements made at lower energy scales in $B$ decays. Within the Standard Model, this value can be reinterpreted as the branching fraction
\begin{equation}{\cal B}(t\rightarrow b\bbar c)=(8\pm4)\times10^{-4}.
\end{equation}

This first measurement of \Vcb using top quark decays  demonstrates the potential to measure \Vcb at hadron colliders, and sets the stage for more precise measurements at the LHC in the future. Such measurements will profit from significant progress in flavour tagging algorithms, larger data samples and extended analysis categories.

% End of text imported from the .//sections/conclusions.tex input file

%
%
%
%
%
\section*{Acknowledgements}
%

%

% The next lines are included from the .//acknowledgements/Acknowledgements.tex input file
%
%

%
%

We thank CERN for the very successful operation of the LHC and its injectors, as well as the support staff at
CERN and at our institutions worldwide without whom ATLAS could not be operated efficiently.

The crucial computing support from all WLCG partners is acknowledged gratefully, in particular from CERN, the ATLAS Tier-1 facilities at TRIUMF/SFU (Canada), NDGF (Denmark, Norway, Sweden), CC-IN2P3 (France), KIT/GridKA (Germany), INFN-CNAF (Italy), NL-T1 (Netherlands), PIC (Spain), RAL (UK) and BNL (USA), the Tier-2 facilities worldwide and large non-WLCG resource providers. Major contributors of computing resources are listed in Ref.~\cite{ATL-SOFT-PUB-2026-001}.

We gratefully acknowledge the support of ANPCyT, Argentina; YerPhI, Armenia; ARC, Australia; BMWFW and FWF, Austria; ANAS, Azerbaijan; CNPq and FAPESP, Brazil; NSERC, NRC and CFI, Canada; CERN; ANID, Chile; CAS, MOST and NSFC, China; Minciencias, Colombia; MEYS CR, Czech Republic; DNRF and DNSRC, Denmark; IN2P3-CNRS and CEA-DRF/IRFU, France; SRNSFG, Georgia; BMFTR, HGF and MPG, Germany; GSRI, Greece; RGC and Hong Kong SAR, China; ICHEP and Academy of Sciences and Humanities, Israel; INFN, Italy; MEXT and JSPS, Japan; CNRST, Morocco; NWO, Netherlands; RCN, Norway; MNiSW, Poland; FCT, Portugal; MNE/IFA, Romania; MSTDI, Serbia; MSSR, Slovakia; ARIS and MVZI, Slovenia; DSI/NRF, South Africa; MICIU/AEI, Spain; SRC and Wallenberg Foundation, Sweden; SERI, SNSF and Cantons of Bern and Geneva, Switzerland; NSTC, Taipei; TENMAK, T\"urkiye; STFC/UKRI, United Kingdom; DOE and NSF, United States of America.

Individual groups and members have received support from BCKDF, CANARIE, CRC and DRAC, Canada; CERN-CZ, FORTE and PRIMUS, Czech Republic; COST, ERC, ERDF, Horizon 2020 and Marie Sk{\l}odowska-Curie Actions, European Union; Investissements d'Avenir Labex, Investissements d'Avenir Idex and ANR, France; DFG and AvH Foundation, Germany; Herakleitos, Thales and Aristeia programmes co-financed by EU-ESF and the Greek NSRF, Greece; BSF-NSF and MINERVA, Israel; NCN and NAWA, Poland; La Caixa Banking Foundation, CERCA and AGAUR programs from Generalitat de Catalunya and PROMETEO and GenT Programmes Generalitat Valenciana, Spain; G\"{o}ran Gustafssons Stiftelse, Sweden; The Royal Society and Leverhulme Trust, United Kingdom; Eric and Wendy Schmidt Fund for Strategic Innovation, United States of America.

In addition, individual members wish to acknowledge support from Chile: Agencia Nacional de Investigaci\'on y Desarrollo (ANID FONDECYT reg. 1230987, FONDECYT 1230812, FONDECYT 1240864, Fondecyt 3240661, Fondecyt Regular 1240721); China: Chinese Ministry of Science and Technology (MOST-2023YFA1605700, MOST-2023YFA1609300), National Natural Science Foundation of China (NSFC 12275265, NSFC-W2543005); Czech Republic: Czech Science Foundation (GACR - 24-11373S), Ministry of Education Youth and Sports (ERC-CZ-LL2327, FORTE CZ.02.01.01/00/22\_008/0004632), PRIMUS Research Programme (PRIMUS/21/SCI/017); EU: H2020 European Research Council (ERC - 101002463); European Union: European Research Council (BARD No. 101116429, ERC - 101219398, ERC - 948254, ERC 101089007), European Regional Development Fund (HE COFUND GA No.101081355, ERDF), Marie Sklodowska-Curie Actions (GAP-101168829); France: Agence Nationale de la Recherche (ANR-21-CE31-0013, ANR-22-EDIR-0002, ANR-24-CE31-0504-01); Germany: Deutsche Forschungsgemeinschaft (DFG - 469666862); China: Research Grants Council (GRF); Italy: Istituto Nazionale di Fisica Nucleare (LHC-MIUR - 28003/2025), Ministero dell'Università e della Ricerca (NextGenEU 153D23001490006 M4C2.1.1, NextGenEU I53D23000820006 M4C2.1.1, NextGenEU I53D23001490006 M4C2.1.1, SOE2024\_0000023); Japan: Japan Society for the Promotion of Science (JSPS KAKENHI  JP25H0063, JSPS KAKENHI JP22H01227, JSPS KAKENHI JP22H04944, JSPS KAKENHI JP22KK0227, JSPS KAKENHI JP24K23939, JSPS KAKENHI JP24KK0251, JSPS KAKENHI JP25H00650, JSPS KAKENHI JP25H01291, JSPS KAKENHI JP25K01011, JSPS KAKENHI JP25K01023); Poland: Polish National Science Centre (NCN 2021/42/E/ST2/00350, NCN OPUS 2023/51/B/ST2/02507, NCN OPUS nr 2022/47/B/ST2/03059, NCN UMO-2019/34/E/ST2/00393, UMO-2022/47/O/ST2/00148, UMO-2023/49/B/ST2/04085, UMO-2023/51/B/ST2/00920, UMO-2024/53/N/ST2/00869); Spain: Agència de Gestió d'Ajuts Universitaris i de Recerca. (AGAUR - 2023 BP 00141), Ministry of Science and Innovation (RYC2019-028510-I, RYC2020-030254-I, RYC2021-031273-I, RYC2022-038164-I), Ministerio de Ciencia, Innovación y Universidades/Agencia Estatal de Investigaci\'on (EU NextGenerationEU (PRTR-C17.I1), PID2022-142604OB-C22); Sweden: Carl Trygger Foundation (Carl Trygger Foundation CTS 22:2312), Swedish Research Council (Swedish Research Council 2023-04654, VR 2021-03651, VR 2022-03845, VR 2022-04683, VR 2023-03403, VR 2024-05451, VR 2025-05940), Knut and Alice Wallenberg Foundation (KAW 2023.0366); Switzerland: Swiss National Science Foundation (SNSF - PCEFP2\_194658); United Kingdom: The Binks Trust, Royal Society (NIF-R1-231091); United States of America: U.S. Department of Energy (ECA DE-AC02-76SF00515), John Templeton Foundation (John Templeton Foundation 63206), Neubauer Family Foundation.

%
%

% End of text imported from the .//acknowledgements/Acknowledgements.tex input file

%
\clearpage
\printbibliography
\clearpage
% ATLAS Collaboration author list
% Reference date of TOPQ-2019-14 is 2026-02-13
% Author list last updated on date 15-MAY-26
% Data extracted on 15-May-2026 for paper reference TOPQ-2019-14
% at 11:08am
 
\begin{flushleft}
\hypersetup{urlcolor=black}
{\Large The ATLAS Collaboration}

\bigskip

\AtlasOrcid[0000-0002-6665-4934]{G.~Aad}$^\textrm{\scriptsize 102}$,
\AtlasOrcid[0000-0001-7616-1554]{E.~Aakvaag}$^\textrm{\scriptsize 17}$,
\AtlasOrcid[0000-0002-5888-2734]{B.~Abbott}$^\textrm{\scriptsize 121}$,
\AtlasOrcid[0000-0002-0287-5869]{S.~Abdelhameed}$^\textrm{\scriptsize 83b}$,
\AtlasOrcid[0000-0002-1002-1652]{K.~Abeling}$^\textrm{\scriptsize 54}$,
\AtlasOrcid[0000-0001-5763-2760]{N.J.~Abicht}$^\textrm{\scriptsize 48}$,
\AtlasOrcid[0000-0002-8496-9294]{S.H.~Abidi}$^\textrm{\scriptsize 30}$,
\AtlasOrcid[0009-0003-6578-220X]{M.~Aboelela}$^\textrm{\scriptsize 44}$,
\AtlasOrcid[0000-0002-9987-2292]{A.~Aboulhorma}$^\textrm{\scriptsize 36e}$,
\AtlasOrcid[0000-0001-5329-6640]{H.~Abramowicz}$^\textrm{\scriptsize 154}$
% ...
\bigskip
\\

% ...
$^{17}$Department for Physics and Technology, University of Bergen, Bergen; Norway.\\
% ...
$^{30}$Physics Department, Brookhaven National Laboratory, Upton NY; United States of America.\\
% ...
$^{36}$$^{(a)}$Facult\'e des Sciences Ain Chock, Universit\'e Hassan II de Casablanca;$^{(b)}$Facult\'{e} des Sciences, Universit\'{e} Ibn-Tofail, K\'{e}nitra;$^{(c)}$Facult\'e des Sciences Semlalia, Universit\'e Cadi Ayyad, LPHEA-Marrakech;$^{(d)}$LPMR, Facult\'e des Sciences, Universit\'e Mohamed Premier, Oujda;$^{(e)}$Facult\'e des sciences, Universit\'e Mohammed V, Rabat;$^{(f)}$Institute of Applied Physics, Mohammed VI Polytechnic University, Ben Guerir; Morocco.\\
% ...
$^{44}$Physics Department, Southern Methodist University, Dallas TX; United States of America.\\
% ...
$^{48}$Fakult\"{a}t Physik, Technische Universit{\"a}t Dortmund, Dortmund; Germany.\\
% ...
$^{54}$II. Physikalisches Institut, Georg-August-Universit\"{a}t G\"ottingen, G\"ottingen; Germany.\\
% ...
$^{83}$$^{(a)}$Khalifa University of Science and Technology, Abu Dhabi;$^{(b)}$New York University Abu Dhabi, Abu Dhabi;$^{(c)}$United Arab Emirates University, Al Ain;$^{(d)}$University of Sharjah, Sharjah; United Arab Emirates.\\
% ...
$^{102}$CPPM, Aix-Marseille Universit\'e, CNRS/IN2P3, Marseille; France.\\
% ...
$^{121}$Homer L. Dodge Department of Physics and Astronomy, University of Oklahoma, Norman OK; United States of America.\\
% ...
$^{154}$Raymond and Beverly Sackler School of Physics and Astronomy, Tel Aviv University, Tel Aviv; Israel.\\
% ...
$^{a}$ Also at Affiliated with an institute formerly covered by a cooperation agreement with CERN.\\
$^{b}$ Also at An-Najah National University, Nablus; Palestine.\\
$^{c}$ Also at Borough of Manhattan Community College, City University of New York, New York NY; United States of America.\\
$^{d}$ Also at Center for Interdisciplinary Research and Innovation (CIRI-AUTH), Thessaloniki; Greece.\\
$^{e}$ Also at Centre of Physics of the Universities of Minho and Porto (CF-UM-UP); Portugal.\\
$^{f}$ Also at CERN, Geneva; Switzerland.\\
% ...

\end{flushleft}

% Created with Glance <Atlas.Glance@cern.ch>

%

%
%

%

%
%
%
%
%

\end{document}